# Decreasing nuclear volume concentrates DNA and enforces transcription factor–chromatin associations during Zebrafish genome activation

Matthias Reisser and J. Christof M. Gebhardt*

Institute of Biophysics, Ulm University, Albert-Einstein-Allee 11, 89081 Ulm, Germany
*Corresponding author, christof.gebhardt@uni-ulm.de

Zygotic genome activation (ZGA), the onset of transcription after initial quiescence, is a major developmental step in many species, which occurs after ten cell divisions in Zebrafish embryos. How transcription factor-chromatin interactions evolve during early development to support ZGA is largely unknown. We established single molecule tracking in live developing Zebrafish embryos using reflected light-sheet microscopy to visualize the general transcription factor TATA-binding protein (TBP), and developed a novel data acquisition and analysis scheme to extract kinetic information during fast cell cycles. The chromatin-bound fraction of TBP increases during early development, compatible with increasing transcriptional activity. By quantifying TBP and DNA concentrations and their binding kinetics, we device a physical model of how the nuclear volume, which decreases during early development, enhances TBP-chromatin associations. Our single molecule data suggest that the shrinking nucleus is a major driving force and timer of ZGA in Zebrafish embryos.

## Introduction

Zebrafish embryos undergo several major morphogenetic transitions during early development, including the mid-blastula transition (MBT) at the 1000-cell stage and gastrulation at dome stage(1). During MBT, the embryo dramatically switches its transcription programme in the maternal-to-zygotic transition, as maternally inherited mRNA is degraded and the zygotic genome is activated (zygotic genome activation, ZGA) (Figure 1a), accompanied by an increase in cell cycle length(1, 2). Until gastrulation, the volume of the animal cap is approximately constant, while individual cells decrease in size after each cell division, accompanied by a decreasing nucleus, but increasing nucleocytoplasmic volume ratio(1, 3).

ZGA is well characterized on the level of mRNA transcripts, for example the relative occurrence of maternal and zygotic transcript levels at different developmental stages(4-6), and epigenetic changes such as the positioning of nucleosomes(7). In contrast, much less is known about changes that proteins initiating transcription such as transcription factors (TFs) or components of the transcription machinery undergo during early development. It is unclear whether TFs are able to bind to DNA in early developmental stages and how DNA accessibility, TF binding to DNA or kinetic properties of TFs change during development. We chose to answer these questions by the example of the general TF TATA-binding protein (TBP). TBP constitutes a representative TF since it is a member of the transcription complex of many genes in Zebrafish(8) that directly binds to DNA.

Binding of TFs to DNA and the assembly of the transcriptional machinery are intrinsically stochastic processes(9-13). Thus, single molecule imaging has evolved as the method of choice to obtain kinetic and quantitative information on TF-DNA interactions. Single molecule imaging has been performed in bacteria(14, 15), individual eukaryotic cells(12, 16-20), large salivary gland cells or cell spheroids(21-23) and whole embryos or adult organisms(24-27). In living Zebrafish, single molecule imaging of YFP-tagged membrane proteins has been performed in epithelial cells at the surface of live Zebrafish using total internal reflection fluorescence (TIRF) microscopy(24).

Here, we established single molecule imaging deep within live growing Zebrafish embryos to study the binding of TBP to chromatin and its regulation during the time course of development embracing ZGA. We considered the parameters contributing to TBP-chromatin interactions, the concentrations of TBP molecules and DNA binding sites and binding reaction rates, and identified the size of the nucleus, which decreases during early development, as major regulator of TBP-chromatin associations. In addition, using TBP as sensor for DNA accessibility, our data suggest that chromatin becomes compacted already during replication.

## Results

### Reflected light-sheet microscopy enables single molecule imaging of chromatin-bound mEos2-TBP in live Zebrafish embryos

To image individual TBP molecules in the nucleus of Zebrafish cells, we adapted reflected light-sheet microscopy (RLSM)(12) to the specific requirements given by the size, shape and medium conditions of live Zebrafish embryos (Figure 1b, Supplementary Figure S1 and Methods). We used a large mirror with rotary mount and a long-range z-stage for sample placement. The reflecting mirror was positioned by remote control, which allowed for user-friendly handling of the RLS microscope. We microinjected mRNA coding for mEos2-TBP and GFP-Lap2$\beta$ as marker of the nuclear envelope into dechorionated fertilized eggs at the 1-cell stage (Figure 1a, Supplementary Figure S2 and S3 and Methods). After photoswitching and exciting mEos2 with 405 nm and 561 nm lasers in a thin plane of a cell nucleus, we could clearly detect the fluorescent signal from individual mEos2-TBP molecules (Figure 1c and Supplementary Movies 1 and 2). As criterion for chromatin binding we required a mEos2-TBP fusion protein to be localized within 0.2 µm$^2$ for at least 100 ms (Methods)(12, 20). A sudden drop of fluorescence to the background level after several frames indicated the single molecule nature of the fluorescence signal (Figure 1c). The signal-to-noise ratio (SNR) decreased from 5 at the surface of the animal cap to 2-3 at a distance up to approx. 60 µm in the embryo (Figure 1d).

### The chromatin-bound fraction of TBP increases during ZGA

Having established single molecule detection in living Zebrafish embryos, we investigated whether binding of TBP to chromatin changes during the first developmental stages embracing ZGA. We quantified the concentration of chromatin-bound mEos2-TBP molecules at successive developmental stages by counting single binding events in the illuminated nuclear volume at constant photoactivation laser power (Figure 2a and Methods). While mEos2-TBP sporadically associated with chromatin in 64-cell embryos, the relative concentration of chromatin-bound mEos2-TBP molecules increased by more than a factor of 25 until the oblong stage. This holds although we excluded the effect of an increasing cell cycle length that by itself increases the number of chromatin-bound mEos2-TBP molecules in this developmental period (Supplementary Figure S4 and Methods). The increase in bound mEos2-TBP concentration was accompanied by an ~6-fold increase in the total nuclear concentration of mEos2-TBP molecules (Figure 2b and Methods). An equal rise of total and bound concentration would be expected from the law of mass action (Supplementary Figure S5). The disproportionality we observe indicates that parameters in addition to the total concentration contribute to regulating the number of chromatin-bound TBP molecules. Nuclear size might be the missing parameter, since we observe an ~4-fold decrease in nuclear size during early development (Figure 2c).

The increase of chromatin-bound mEos2-TBP might not reflect the evolution of endogenous TBP-chromatin associations if both species exhibit different nuclear concentrations, for example due to differences in translational kinetics. However, in *Xenopus*, a comparable increase of endogenous TBP expression levels has been observed before ZGA(28). To be able to draw conclusions also valid for endogenous TBP, we determined the fraction of chromatin-bound TBP molecules, a quantity independent of molecule concentration (Figure 2d and Methods). As expected from the differential behaviour of bound and total mEos2-TBP concentrations, the fraction of bound TBP molecules increased ~4.5-fold between the 64-cell and the oblong stage. Again, this manifests a deviation from the law of mass action, which would predict a constant bound fraction, and indicates a missing parameter for regulating TBP-chromatin associations.

**TBP dissociates from DNA with two distinct dissociation rate constants**

To understand to which extent the parameters entering the law of mass action, concentration and binding rate constant of TBP and concentration of DNA binding sites (Supplementary Figure S5), contribute to regulating chromatin binding of TBP, we next characterized the dissociation rate constant of TBP from chromatin. We performed time-lapse imaging with different dark times(12) of bound mEos2-TBP molecules in oblong embryos, where a high fraction of bound mEo2-TBP facilitated these experiments, monitored the fluorescent 'on' times of mEos2-TBP and extracted the dissociation rate constants using a bi-exponential decay model (Figure 2e, Supplementary Figure S6 and Methods). A fraction of TBP molecules interacted transiently with chromatin with an average residence time of (0.26 ± 0.03) s, while the remaining fraction of molecules bound more stable to chromatin with a long residence time of (6.83 ± 0.74) s. Our results are consistent with previous estimates of the TBP residence time on the order of seconds in living yeast(29). A similar biphasic kinetic behaviour has previously been observed for various other TFs and was identified with transient binding to unspecific sequences during a search process until binding to specific target sequences on chromatin(30-37).

The full assessment of dissociation rate constants requires measurements at many different time-lapse conditions(38), which is challenging during the fast cell cycles of the early embryo. To assess the TBP-chromatin interaction kinetics on faster time scales, we thus developed a novel illumination scheme, interlaced time-lapse microscopy (ITM). In this illumination scheme, two subsequent image acquisitions are followed by a long dark time, and only one dark time condition is sufficient to obtain quantitative information on the fraction of specifically bound molecules (Supplementary Figure S7 and Supplementary Information). We used ITM to quantify the fraction of specific mEos2-TBP-chromatin interactions between the 64-cell and the oblong stage and found this value to be constant to good approximation (Figure 2f). Since ITM is sensitive to a change in length of binding interactions (Supplementary Information), a constant value suggests that the dissociation rate constants of TBP do not change during early development. This result also holds for endogenous TBP, as the fraction of specifically bound molecules is independent of concentration.

**Nuclear size regulates TBP-chromatin associations**

For quantitative analysis, we focused on the fraction of bound TBP molecules, since this ratio is independent of the total amount of visible mEos2-TBP molecules and thus allows maximizing this number by adjusting the photoactivation laser power for high-throuput data acquisition (Figure 3a and Methods). We corrected the fraction of bound molecules for photobleaching of mEos2, using parameters from our kinetic measurements (Methods). We further assumed that the concentration of accessible DNA binding sites was constant during ZGA, based on the observation that nucleosomes occupy a constant fraction of ~80% of the DNA at early developmental stages in Zebrafish embryos(7). We applied our minimal kinetic binding model in a fit to the fraction of bound TBP molecules, considering the measured parameters of relative TBP concentration, the TBP dissociation constant and nuclear size (Figure 3a and Supplementary Information). The model very well reproduced the overall increase with low residuals (Supplementary Figure S8). From the fit, we extracted the number of accessible DNA binding sites, the only free parameter of our model (Figure 3b).

Our model predicts that a higher concentration of accessible DNA binding sites will increase the fraction of bound TBP molecules. To test this prediction, we determined the fraction of bound TBP molecules in embryos in the presence of Trichostatin A (TSA) between the 64-cell and the oblong stage (Supplementary Figure S9 and Methods). As expected, the fraction of bound TBP molecules increased considerably, accompanied by an increase of accessible DNA binding sites (Figure 3b).

The quantitative analysis of our data suggests a model for regulating TBP-chromatin interactions during ZGA (Figure 3c), inspired by previous reflections on the importance of nuclear size and concentration(39, 40): TBP searches for specific target sequences associated with long residence time while scanning chromatin with transient interactions. While the overall ratio of specific to unspecific associations does not change during ZGA, the decreasing nuclear size, which influences protein and DNA concentrations, ensures that increasingly higher fractions of TBP are associated with chromatin during this important developmental period in the early embryo. The starting value of this fraction is set by the concentration of accessible DNA binding sites (Figure 3a). In contrast, the TBP concentration determines the absolute number of TBP molecules associated to specific target sites (Figure 3d). Similarly, the specific residence time of TBP influences the amount of specifically bound TBP molecules (Figure 3d).

**Chromatin is compacted during replication**

Our quantitative measurement of the fraction of bound TBP molecules proved to be a sensitive method to determine the number of accessible DNA binding sites for TBP. We thus used TBP as a sensor to assess whether the number of accessible DNA binding sites changes during replication. We split the single molecule tracking data within each nucleus into events early after cell division (pre-replication) and shortly before the next cell division (post-replication) and determined the fraction of bound TBP molecules (Figure 4a and Methods). A larger fraction of TBP molecules bound to chromatin up to the 1000 cell stage in pre-replication compared to post-replication nuclei. At the same time, the size of the nucleus increased during replication (Supplementary Figure S10). Overall, the number of accessible DNA binding sites did not change significantly during replication (Figure 4b). This result suggests that DNA becomes compacted during replication such that sites accessible for TBP molecules approximately halve.

## Discussion

**Applicability of reflected light sheet microscopy to single molecule imaging in Zebrafish embryos**

In principle our modifications to the RLS microscope allow movement of the light-sheet focal plane to any position within the embryo. In practice, single molecule imaging is limited in height above the sample surface by the working distance of the detection objective, which in our case was ~300 µm. Single molecule imaging in the embryo was limited to a depth of approx. 60 µm from the surface of the animal cap. This limit is due to absorption and scattering of fluorescent light within the tissue of the embryo(41). Once the cell radius decreased below this threshold, single molecule imaging in live growing Zebrafish embryos became straightforward using RLS microscopy. Given the relatively large size of Zebrafish embryos compared to other model systems, RLS microscopy might very well be suited to image single fluorescent molecules also in other model systems such as *C. elegans*, *D. melanogaster*(26), *in vitro* grown mouse blastocysts or organoids.

**A concentration model for the regulation of TF – chromatin associations**

To model the chromatin binding properties of TBP, in particular the increase in the bound fraction of TBP during early development, we considered all parameters of the law of mass action contributing to TBP-DNA associations: the concentrations of TBP and DNA binding sites and the kinetic rate constants of TBP binding to unspecific and specific DNA sequences. This quantitative analysis of our single molecule data identified the size of the nucleus, which decreases during early cell division cycles and thereby influences the concentrations of proteins and DNA, as the driver of the increasing bound TBP fraction. Since changes in nuclear size globally affect the concentrations of all biomolecules in the nucleus, it is likely that other DNA-binding factors are modulated similar to TBP in their binding behaviour to DNA. The decrease of the nuclear size thus might play a universal role in controlling the fraction of DNA-bound molecules, in particular TFs, during the first developmental stages in Zebrafish embryos. While the basis of nuclear size regulation is not yet well understood, scaling between cell size and nuclear size appear to be a widespread phenomenon and several principle mechanisms and possible regulators have been suggested(42, 43). Nuclear size may be regulated by a limiting cytoplasmic pool of building subunits(39) such as components of the nuclear envelope(44) or the nuclear pore complex(45), by nuclear import(46) or a force balance between cytoskeletal and nuclear components(47).

In contrast to the bound fraction, the absolute number of specifically bound TBP molecules depends on the amount of TBP in the nucleus, in addition to nuclear size influencing the binding site concentration. As binding of TBP is an essential step in transcription of many genes in Zebrafish(8), it is this parameter that should be correlated to transcriptional activity. Questions related to the regulation of transcriptional activity in the embryo thus need to address the number of maternally inherited TBP molecules and mRNAs initially present in the Zebrafish embryo(48) as well as the kinetics of TBP translation in addition to nuclear size(28).

In the early embryo, nucleosome occupancy of DNA has been observed to be unbiased and mostly free of the regular positioning seen in transcriptionally active cells(7). Thus, unspecific DNA sequences and specific TBP target sites will be accessible to TBP without bias. Consistent with this notion, we measured a ratio of specifically bound to all bound TBP molecules that is approximately constant between the 64-cell and the oblong stage. The initial fraction of specifically bound TBP molecules then results from the DNA-encoded ratio of specific to unspecific sequences and the relative values of specific and unspecific dissociation rate constants of TBP. To which degree pioneering TFs(49-51) acting at early developmental stages or chromatin remodelers(52, 53), that have been shown to be essential in mouse embryogenesis(54), might modulate the accessibility to TBP of DNA and thereby contribute to the association of the transcription machinery during development will be important to solve in the future.

Replication did not increase the concentration of accessible DNA binding sites. This is consistent with observations that transcriptional activity does not increase considerably after replication in several species(55, 56) and might be linked to chromosome compaction starting during replication(57). In mitosis, the concentration of accessible DNA binding sites is minimized by chromatin compaction, and our model trivially predicts a reduction of DNA-bound TFs, which is observed for many TFs(58).

**The concentration model and temporal regulation of ZGA**

The increase we observe for the bound TBP fraction and for the number of specifically bound TBP molecules coincides in time with the onset of transcription at ZGA in the early Zebrafish embryo(59). Due to the important role of TBP in transcription, this correlation likely reflects a causal relation. Since we observe a gradual increase in specific TBP-DNA interactions starting from the 64-cell stage onwards, it is conceivable that zygotic transcription, too, starts to increase at this early developmental stage. Our data thus suggest that ZGA is a gradual rather than a sudden process. Consistent with this interpretation, a gradual increase of transcription levels and the number of transcribed genes has been observed starting from the 64-cell stage onward(4-6).

Several mechanisms considering TF binding have been suggested to underlie the activation of zygotic transcription and regulate its timing(60, 61). One model suggests that exponentially increasing amounts of DNA within the roughly constant volume of the early *Xenopus* embryo titrate out a repressor(62, 63). Experiments in which a plasmid encoding a reporter gene was injected into *Xenopus* embryos suggested inhibition of TBP binding to DNA by competitive assembly of nucleosomes(64). Further, a deficiency of components of the transcription machinery has been shown to contribute to the absence of transcription before genome activation(28). Recent experiments in Zebrafish suggest that a competition for binding to DNA between histone octamers and TFs regulates the onset of transcription during genome activation(3).

For our measurements, the concentration model including changes to the nuclear size is sufficient to explain the increase in the bound fraction of TBP. Naturally, every TF in the nucleus and their associated DNA target sites will undergo similar changes in concentration and shift towards the bound state as the size of the nucleus decreases. The concentration model, however, does not exclude or disprove repressor titration or competition models and they most probably coexist. It is unlikely that the competition model does not apply to TBP binding, given that we observe short DNA residence times of TBP comparable to other TFs, and competition might account for a small modulation of accessible DNA binding sites within our error intervals. The experiments that led to previous models of ZGA timing included injecting high amounts of excess DNA or histones such that a small repressive or competitive effect on TF-chromatin binding might have become amplified until observable. Since concentrations and kinetic properties were not considered to full extend, the potentially dominant impact of decreasing nuclear size on TF-chromatin associations during ZGA might have been underestimated. To which extent nuclear size, the concentration of accessible DNA binding sites and binding competition between histones and TBP contribute to the temporal regulation of ZGA is subject to future studies.

Some genes precede the general onset of transcription in 1000-cell Zebrafish embryos by several cell cycles(4-6). Our concentration model suggests two major mechanisms to achieve efficient early DNA association of a TF and thus gene-specific transcription activation: by increasing the concentration of this protein or by a longer DNA residence time. Consistent with this suggestion, the DNA residence times of pioneering TFs such as Sox2 and Oct4 have been observed in embryonic stem cells to be longer than the DNA residence time of TBP determined here(31). According to our model, such a residence time would allow for substantial association to DNA already in the 128-cell stage. Whether the embryo ensures early transcription of a gene by regulating the DNA residence time, the concentration of TFs or yet other mechanisms will probably be gene-specific.

While differing in the details, other species including *Drosophila*, *Xenopus,* sea urchins and *C.elegans* share common characteristics with Zebrafish during early stages of development. Their embryonic genomes are transcriptionally silent for several cell division cycles(59). In these species, the size of individual nuclei decreases considerably during the initial phase of rapid cell divisions before ZGA(65-68). Such a decrease is also observed in mammals(69). In *Xenopus*, an effect of nuclear size on the timing of ZGA has been shown(70). It is thus very likely that a concentration model similar to the one we find responsible for timing TF binding and thus transcription onset in Zebrafish embryos also applies to other species.

**Conclusion**

We have performed single molecule tracking experiments in live developing Zebrafish embryos to investigate the binding to DNA of the general TF TBP at different developmental stages. Our experiments and mathematical modeling reveal that a simple mechanism of decreasing nuclear volume is able to control the associated fraction of TBP and probably other TFs to DNA and thus might be responsible for the important developmental step of zygotic genome activation. Since mainly protein concentrations determine how many TF molecules will actually be bound to specific DNA target sites at any given developmental stage and thus might initiate transcription, an important role is assigned to the control of protein numbers, both within the unfertilized egg as well as during the first developmental stages. The compelling nature of our concentration model lies in its simplicity and potential generality to describe associations of proteins such as TFs to DNA, while providing the molecular and kinetic foundations underlying the process of zygotic genome activation.

**Author contributions**

J.C.M.G. and M.R. designed experiments. M.R. constructed the reflected light sheet microscope. M.R. performed experiments. M.R. and J.C.M.G. analyzed data. J.C.M.G. and M.R. wrote the manuscript.

**Acknowledgements**

We thank Nadine L. Vastenhouw, Shai R. Joseph and Máté Pálfy (Max Planck Institute of Molecular Cell Biology and Genetics, Dresden, Germany) for the introduction to Zebrafish techniques, the mEos2-TBP and GFP-Lap2β mRNA (GFP-Lap2β originates from Marija Matejcic, Caren Norden lab, Max Planck Institute of Molecular Cell Biology and Genetics, Dresden, Germany), discussions and comments to an early version of the manuscript and Gilbert Weidinger and Christopher Jahn (Institute of Biochemistry and Molecular Biology, Ulm University, Ulm, Germany) for technical support and support with their Zebrafish facility. The work was in part funded by the German Research Foundation (No. GE 2631/1-1 to J.C.M.G.), the European Research Council (ERC) under the European Union's Horizon 2020 Research and Innovation Program (No. 637987 ChromArch to J.C.M.G.) and the German Academic Scholarship Foundation (to M.R.).

**Methods**

**Reflected light-sheet (RLS) microscopy**

We modified the reflected light-sheet microscope based on a previously published version(12).

*Illumination optics*

Laser light of 405 nm (Laser MLD, 200mW, Cobolt, Solna, Sweden), 488 nm (IBEAM-SMART-488-S-HP, 200 mW, Toptica, Graefelfing, Munich, Germany) and 561 nm (Jive 300 mW, Cobolt) was expanded to 1.2 mm beam diameter by Kepler telescopes with lenses of a focal length of 75 mm and 100 mm for the 488 nm laser and of 60 mm and 150 mm for the other lasers. Afterwards it was combined using dichroic mirrors (F48-486, F43-404, AHF, Tuebingen, Germany) and selectively filtered by an AOTF (AOTFnC-400.650-TN, AA Optoelectronics, Orsay, France). Subsequently, light was coupled into a single-mode fiber (S405, Thorlabs, Dachau, Germany) by a fiber-coupler (60FC-4-RGBV11-47, Schäfter+Kirchhoff, Hamburg, Germany). After the fiber, the beam was expanded by a fiber collimator (60FC-T-4-RGBV42-47, Schäfter+Kirchhoff) to a diameter of 1.5 cm and unilaterally focused by a cylindrical lens (f=15 cm, ACY254-150-A, Thorlabs) into the back-focal plane of a water dipping objective ( 40x 0.8 NA HCX Apo L W, Leica, Wetzlar, Germany). The effective NA was reduced by an iris after the cylindrical lens and setting a beam diameter of 1.5 mm before the water dipping objective. The light sheet formed by the objective was reflected by the chip of an AFM cantilever (HYDRA2R-100N-TL-20, AppNano, Mountain View, CA) that was coated with 4 nm Titanium and subsequently with 40 nm Aluminum by physical vapor deposition (PVD). The resulting light sheet had a thickness of approximately 3 µm. Bright field illumination was achieved by a LED (720nm, HCA1 H720, 70 mW, Roithner, Vienna, Austria) whose light was overlaid with the laser-light before the illumination objective by a dichroic mirror (F38-635, AHF).

*Mechanical parts*

The cantilever used as micro-mirror was mounted on a xyz-micro-positioning stage (Witech, Ulm, Germany) to enable precise remote-controlled positioning of the cantilever. The whole composition of bright-field illumination, fiber-collimation, cylindrical lens and illumination objective was mounted onto a rotatable stage (XYR1/M, Thorlabs) positioned on a tripod which was mounted on a commercial microscope body (TiE, Nikon, Duesseldorf, Germany). The sample dish was mounted on a piezo z-stage with 500 µm travel (Nano-ZL 500, Mad City Labs, Madison, WI).

*Detection optics*

Fluorescence light was collected with a water-immersion objective (60x 1.20 NA Plan Apo VC W, NIKON) filtered by a dichroic mirror (F73-866 / F58-533, AHF), an emission filter (F72-866 / F57-532, AHF) and a notch filter (F40-072 / F40-513, AHF). before being post-magnified (1.5x) and detected by an EM-CCD Camera (iXon Ultra DU 897U, Andor, Belfast, UK).

*Electronic control*

All electronic parts of the setup were controlled using NIS Elements software (Nikon) and a NIDAQ data acquisition card (National Instruments, Austin, TX). Macros were written in a C-like NIS Elements macro language.

**Sample preparation**

We dechorionated wild-type AB Zebrafish embryos directly after fertilization and synchronized them by visual selection. We injected 60pg mEOS2-TBP mRNA (gift from Nadine L. Vastenhouw, Max Planck Institute of Molecular Cell Biology and Genetics, Dresden, Germany) and 9pg GFP-Lap2β mRNA (gift from Nadine L. Vastenhouw and originating from Marija Matejcic from the Caren Norden lab, Max Planck Institute of Molecular Cell Biology and Genetics, Dresden, Germany) in 1 nl into the animal cap at the 1-cell stage to visualize the transcription factor and the outline of the nucleus(71). For TSA injected fish, 1.16 pg TSA in 3 nl were also injected at the 1-cell stage into the yolk. Embryos developed to the 32-cell stage at 28°C. At the 32-cell stage, we mounted embryos on a glass bottom dish (Delta T, Bioptechs, Butler, PA) at room temperature. During fluorescence imaging, we counted cell cycles based on the breakup and re-formation of the nuclear envelope labeled with GFP-Lap2β. The embryos developed to sphere stage within 4.5 hours, after which fluorescence imaging was stopped.

**Data acquisition**

*Time-lapse illumination*

Generally, we took videos of mEos2 fluorescence after an initial period of photo-activation. For concentration-sensitive measurements, we kept the photo-activation period constant for all nuclei of an embryo at levels resulting in a density of activated mEos2 molecules suitable for single molecule detection. We adjusted the photo-activation period from 0.5-5 s depending on the concentration of mEos2 fusion proteins in the nuclei in case of concentration-insensitive measurements. We set illumination time $\tau_{on}$ with 561nm laser light to 250 ms (mEos2-TBP). Dark times in between illumination times were varied depending on the time-lapse condition.

*Interlaced time-lapse illumination*

As described in detail in Supplementary Information section 2, we implemented a new illumination pattern called interlaced-timelapse microscopy (ITM). Two successive frames were taken during a total illumination time of 330 ms followed by a dark time of 750 ms. For determination of concentration-sensitive values such as nuclear concentration and number of bound molecules, the duration of the photo-activation period was set to 1 s at all developmental stages. We adjusted the photo-activation period from 0.5-5 s depending on the concentration of mEos2 fusion proteins in the nuclei in case of concentration-insensitive measurements.

## Data analysis

*Particle tracking*

We performed data analysis following the procedures published in (12). In brief, bright pixels where identified as candidates for fluorescent molecules, if their grey value was above the mean plus 3.5 standard deviations of the whole image. We fitted a 2D-Gaussian curve to the surrounding area of the pixels. The centre determined by the fitting function was taken as position of the fluorescent emitter. We connected fluorescent spots to tracks according to distance criteria. A spot that that was localized within 0.2 µm² for at least 100 ms in time-lapse microscopy was identified as a bound molecule (12, 20). The effective pixel size in the specimen was measured to be 166 nm. For these steps of analysis, programs written in MATLAB (MathWorks, Natick, MA) were used. Subsequent analysis steps were written in Python 2.7.9.

*Determination of nuclear concentrations of TBP*

To quantify nuclear concentrations of mEos2 fusion proteins we photo-activated the sample several times per cell cycle with a lag time of approx. 2 minutes to allow for sufficient recovery of unactivated molecules. We analysed only the first frame after photo-activation.

*Determination of DNA residence times*

To extract molecular DNA-residence times we implemented a global fitting approach as described in (12) in Python. In brief, the numbers of frames a molecule was detected at the same position (fluorescent 'on' times) were binned to histograms and a double-exponential distribution (equation 1) was globally fitted to the error-weighted histograms using a Levenberg-Marquardt least squares algorithm.

$$f_T(t) = A \cdot \left[ B \cdot \left( k_b \cdot \left( \frac{\tau_{on}}{\tau_{tl}} \right) + k_1 \right) \cdot \exp\left( -\left[ k_b \cdot \left( \frac{\tau_{on}}{\tau_{tl}} \right) + k_1 \right] \cdot t \right) + (1 - B) \cdot \left( k_b \cdot \left( \frac{\tau_{on}}{\tau_{tl}} \right) + k_2 \right) \right.$$
$$\left. \cdot \exp\left( -\left[ k_b \cdot \left( \frac{\tau_{on}}{\tau_{tl}} \right) + k_2 \right] \cdot t \right) \right] \quad (1)$$

Parameters optimized were the bleaching rate $k_b$, the off-rates of the bound TBP molecules $k_1$ and $k_2$, the fraction of long and short bound molecules B and 1 − B and the overall amplitude A. The parameters $\tau_{on}$ on and the time-lapse time $\tau_{tl}$ (including illumination time and dark time) were pre-set and not subject to optimization. The stopping criterion was set to a relative change of $10^{-7}$ in the sum of error-squares. Reduced $\chi^2$ - values were compared between a double- and a single exponential model. For the fit of mEos2-TBP fluorescent 'on' times reduced $\chi^2$ -values of 0.010 and 0.015 were found for the double- and the single-exponential model respectively. These values strongly prefer the double-exponential model over the single exponential one.

*Determination of the fraction of bound molecules*

In interlaced time-lapse microscopy (ITM), successive localizations of mEos2-TBP within 166 nm were classified by their fluorescent 'on' time: Localizations surviving at least one dark time period were classified as long-binding events while events appearing in at least two frames were classified as all binding events (long or short). We counted events for each stage in each embryo. Errors and means of these numbers between sets of embryos were determined by a bootstrapping procedure where a set of 900 random subsets, each containing 80% of the measured data, was analysed. The ratio of long to all binding events R is not exactly equivalent to the ratio of stable to all bound molecules B as bleaching and unbinding would lead to counting of stable binding events to the class of transient binding events. Thus, a correction formula was derived in mathematical detail in Supplementary Information section 3.1. Dividing the number of bound molecules by the number of molecules that are detected at least in one frame yields the ratio F. This ratio can also be underestimating the true proportion of bound to freely diffusing molecules D. Thus, a correction formula was derived in mathematical detail in Supplementary Information section 3.1. Examples for the correction procedure are given in Supplementary Information section 3.1. The concentration model derived in Supplementary Information section 3.2 was fitted to the data using a Levenberg-Marquardt least squares algorithm in Python. The stopping criterion was set to a relative change of $10^{-7}$ in the sum of error-squares.

Figures:

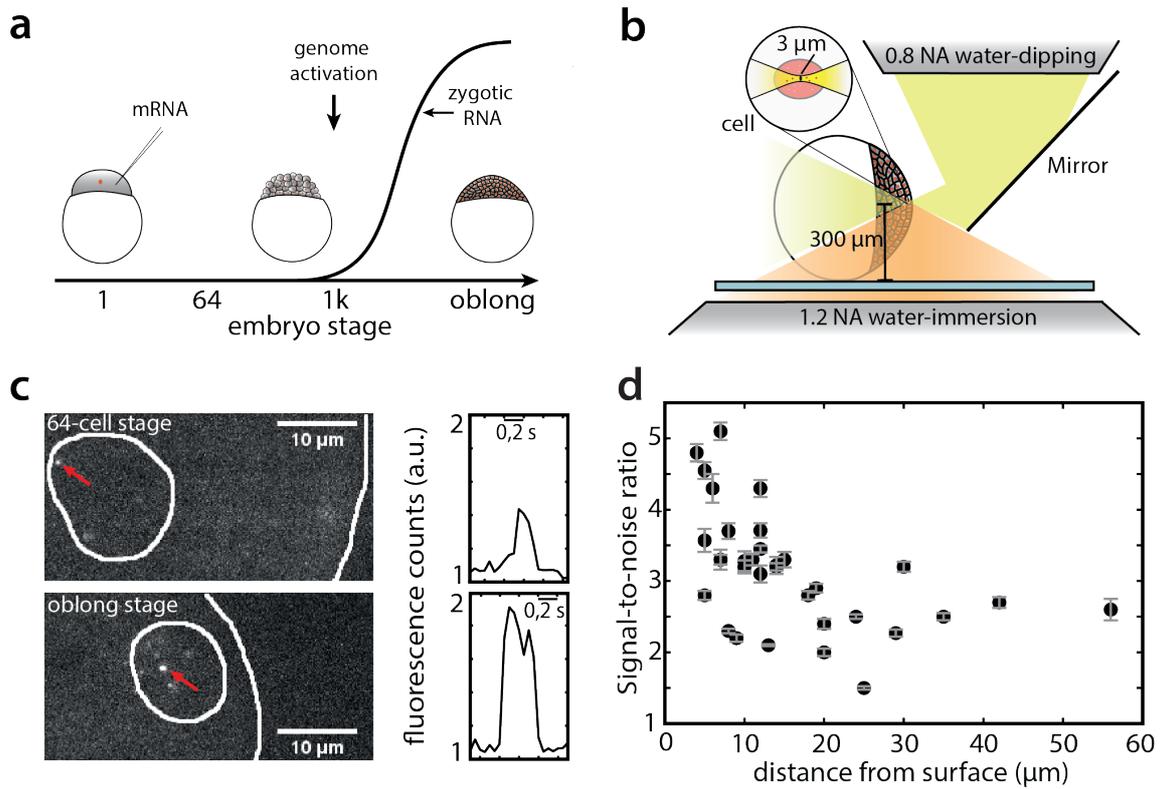

**Figure 1.** Single molecule imaging in live Zebrafish embryos by reflected light-sheet microscopy. (**a**) Sketch of zygotic mRNA levels as a function of embryo stage. *Inset:* schemes of embryos at different developmental stages. *Left:* mRNA injection into the animal cap of 1-cell stage embryos. (**b**) Sketch of reflected light-sheet microscopy of a Zebrafish embryo. *Inset:* close-up view of a single cell within the embryo. (**c**) *Left:* fluorescence images of mEos2-TBP in a nucleus of a 64-cell stage embryo and an oblong stage embryo. The surface of the animal cap and the outline of the nucleus are indicated (white lines). Red arrows point to single mEos2-TBP molecules. *Right:* time traces of mEos2-fluorescence of the molecules indicated in the left panels. (**d**) Signal-to-noise ratio of the single molecule fluorescence signal of nuclear mEos2-TBP as a function of distance of the nucleus from the surface of the animal cap. Values are calculated from all detected molecules within a nucleus that is centred at the given position and represented as mean ± sem.

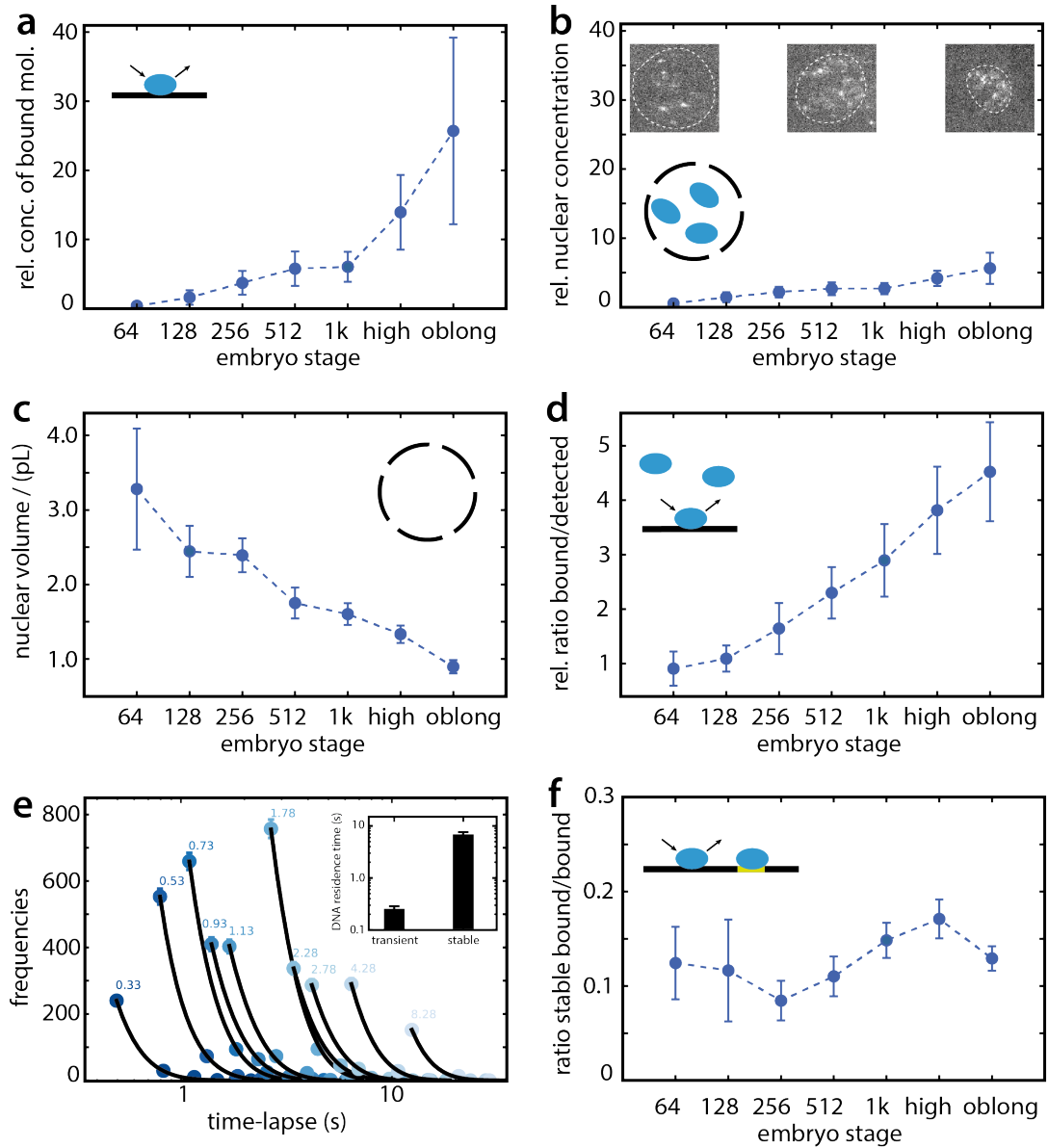

**Figure 2.** The fraction of chromatin-bound TBP increases during early development, while the fraction of specifically bound TBP stays constant. (**a**) Relative concentration of chromatin-bound mEos2-TBP as function of developmental stage (1955 molecules in 6 embryos). (**b**) Relative concentration of nuclear mEos2-TBP as function of developmental stage (27803 molecules in 6 embryos). *Insets:* fluorescent images of mEos2-TBP at different developmental stages. The outline of the nucleus is indicated (dashed white lines). (**c**) Nuclear volume as function of developmental stage (4 embryos). (**d**) Relative fraction of chromatin-bound TBP as function of developmental stage (27803 molecules in 6 embryos). (**e**) Histograms of fluorescent 'on' times at different time-lapse conditions of mEos2-TBP in nuclei of oblong embryos. Data is represented as mean ± sd. Lines: global fit of a bi-exponential decay model (Equation 1 in Methods) (4780 molecules in 15 embryos). *Inset:* DNA residence times of mEos2-TBP. Errors are the sd from the fit. (**f**) Fraction of specifically bound TBP as function of developmental stage (31330 molecules in 11 embryos).

Relative values are normalized to the average value of the 64-cell and 128-cell stages. Data is represented as mean ± sem if not indicated otherwise.

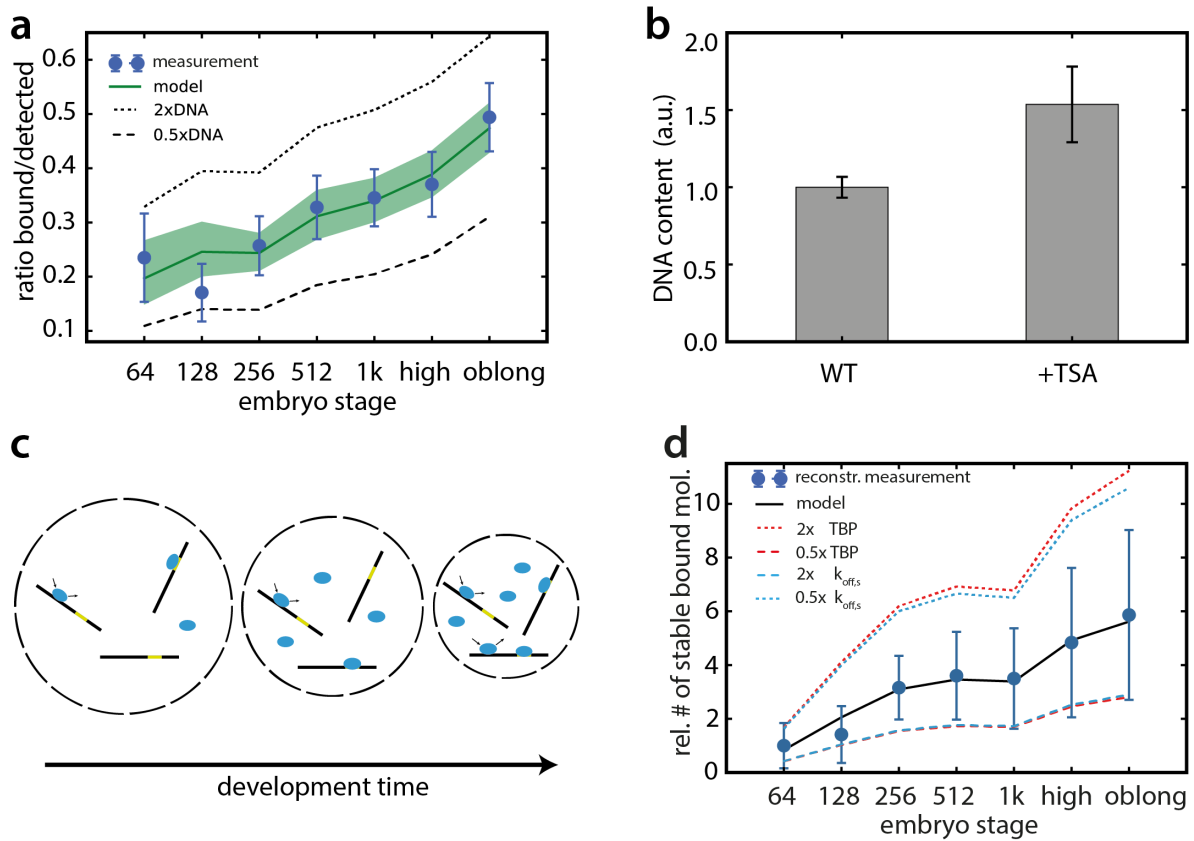

**Figure 3.** The concentration model for TBP-chromatin associations. (**a**) Absolute fraction of chromatin-bound TBP as function of developmental stage (blue spheres), fraction obtained with the concentration model (green line, green shade represents the error interval) and fraction predicted if accessible DNA were doubled (dotted black line) or halved (dashed black line) (534612 molecules in 11 embryos). (**b**) Comparison of DNA sites accessible to TBP between wild type embryos and embryos in the presence of TSA (11 and 3 embryos). (**c**) Scheme of the concentration model. During early embryo development the size of individual nuclei decreases, thereby increasing the concentration of TBP and DNA. This results in an increase in the DNA-bound fraction of TBP. (**d**) Relative number of TBP molecules bound to a specific DNA target sequence as function of developmental stage, normalized to the average value of the 64-cell and 128-cell stages (blue spheres), number obtained with the concentration model (black line) and number predicted if concentration of TBP were doubled (dotted red line) or halved (dashed red line) or if dissociation rate constant of TBP were doubled (dashed blue line) or halved (dotted blue line) (534612 molecules in 11 embryos). Data is represented as mean ± sem.

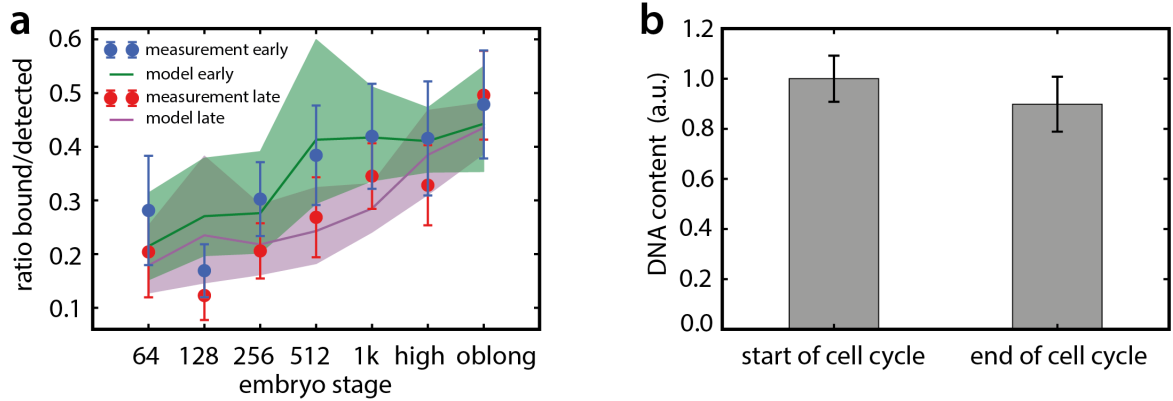

**Figure 4.** Chromatin compacts during replication. (**a**) Absolute fraction of chromatin-bound TBP as function of developmental stage at start of cell cycle (blue spheres) and at end of cell cycle (red spheres) and fraction obtained with the concentration model at start of cell cycle (green line) and at end of cell cycle (red line). Shades represent the error interval (52504 molecules (start of cell cycle) and 57991 molecules (end of cell cycle) in 11 embryos). (**b**) Comparison of DNA sites accessible to TBP between nuclei at start of cell cycle and at end of cell cycle (11 embryos). Data are represented as mean ± sem.

# Supplementary Information

**Decreasing nuclear volume concentrates DNA and enforces TBP-chromatin associations during Zebrafish genome activation**

Matthias Reisser and J. Christof M. Gebhardt[*]

Institute of Biophysics, Ulm University, Albert-Einstein-Allee 11, 89081 Ulm, Germany
[*]Corresponding author `christof.gebhardt@uni-ulm.de`

# Contents

# 1 Supplementary Figures

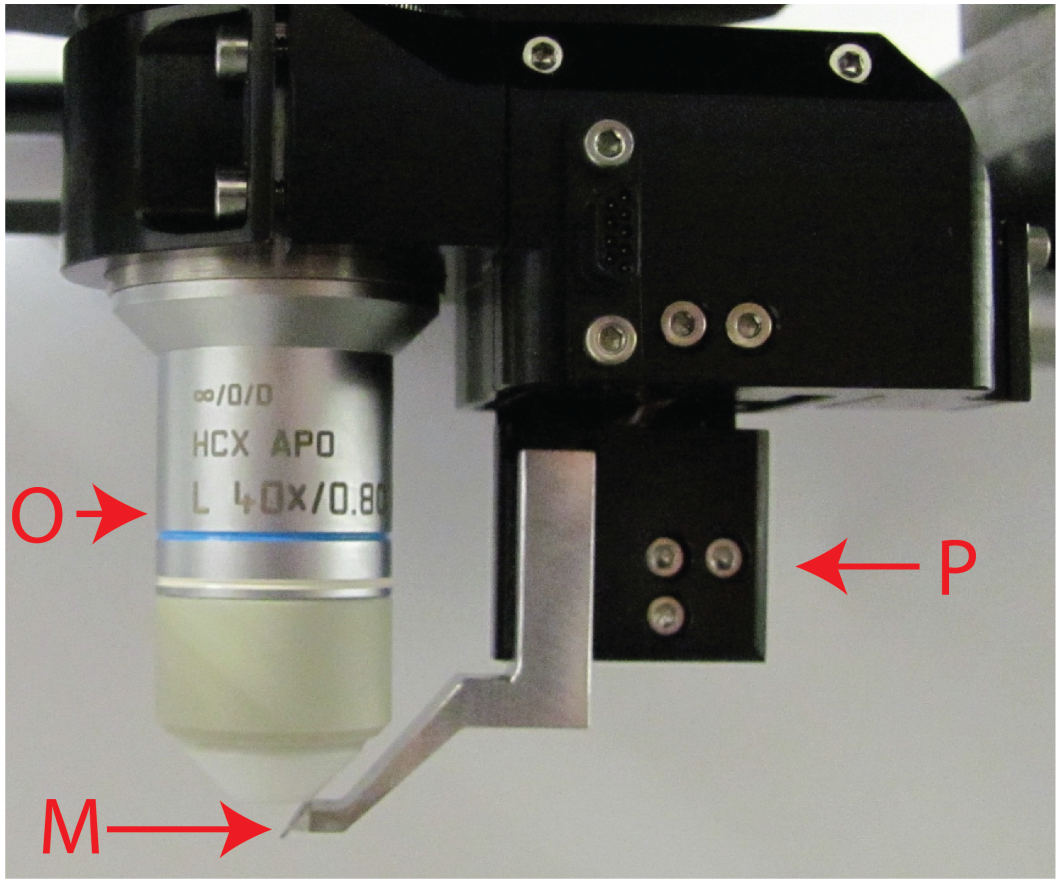

**Supplementary figure S1**
**Illumination unit of the RLS-microscope.** **O** water dipping objective, **M** micro-mirror, **P** x-y-z micro positioning stage

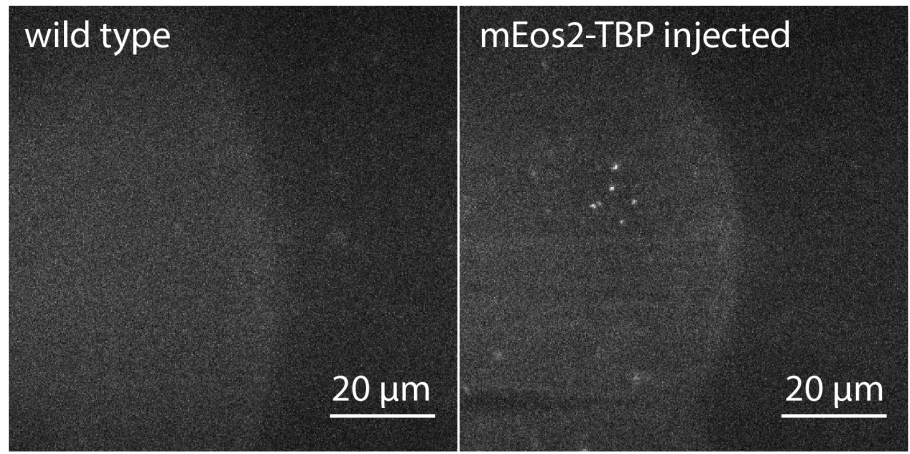

**Supplementary figure S2**
**Comparison of wild type and mEos2-TBP injected embryos**. Wild type embryo and mEos2-TBP injected embryos under illumination with 561 nm laser light after photoactivation with 405 nm laser light. The wild type embryo does not show fluorescent spots. For both embryos at 128-cell stage exposure time was set to 50ms.

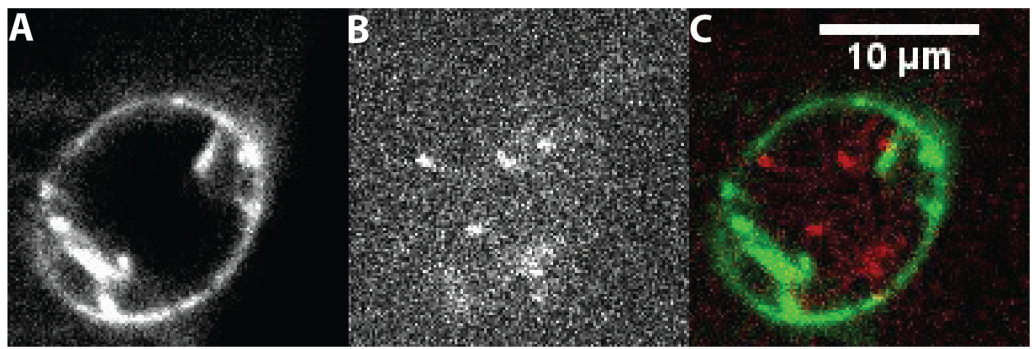

**Supplementary figure S3**
**Visualization of mEos2-TBP in the nucleus.** (**A**) To localize the nucleus inside the cell, GFP-Lap2$\beta$ was excited with 488 nm laser light and fluorescence was detected on an EMCCD camera with 50 ms integration time. (**B**) To localize the TBP molecules inside the cell, mEos2-TBP was excited with 561 nm laser light after photoactivation with 405 nm laser light and fluorescence was detected on an EMCCD camera with 50 ms integration time. (**C**) Merged images A+B in false-colors with separately adjusted contrast for better visibility

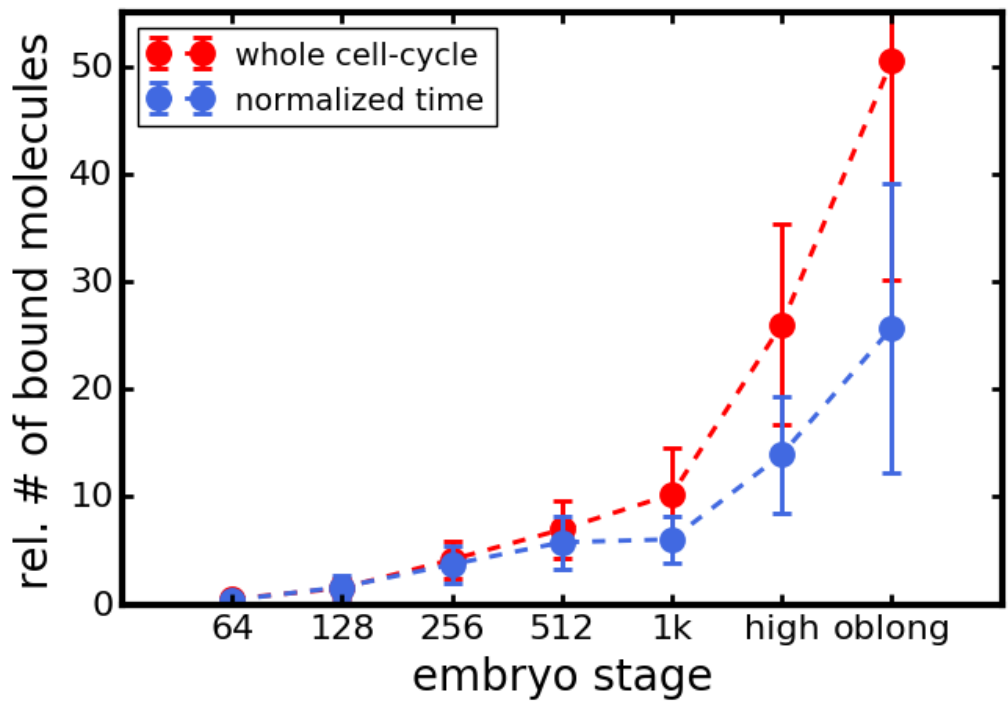

**Supplementary figure S4**
**Longer cell cycles lead to more binding events** The cumulated number of observed DNA-binding molecules per unit volume during the cell cycle (red curve, so called 'dose') increases more than the number of observed DNA-binding molecules per unit volume and unit time (blue curve, so called 'concentration') due to the lengthening of the cell cycle. The curves are each normalized to the average value of the 64-cell and the 128-cell stage. (7418 molecules in 6 embryos (whole cell cycle) and 1955 molecules in 6 embryos (normalized time) ).

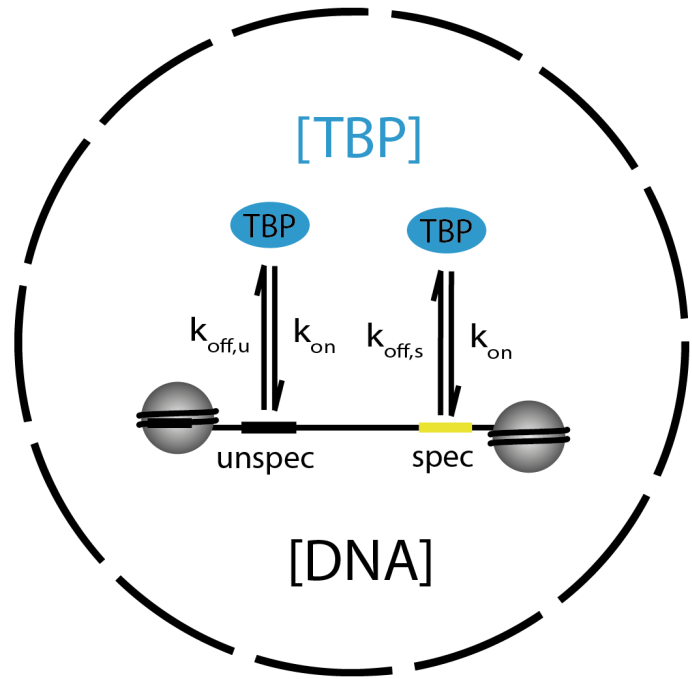

**Supplementary figure S5**
**Kinetic scheme of the concentration model** Following the law of mass action, the concentration of TBP-DNA complexes depends on the concentration of TBP, [TBP], the concentration of DNA, [DNA], and the on-rate $k_{on}$ as well as on the off-rate $k_{off,u}$ from unspecific sequences and $k_{off,s}$ from specific target sites. Concentration of TBP and DNA is strongly influenced by the size of the nucleus.

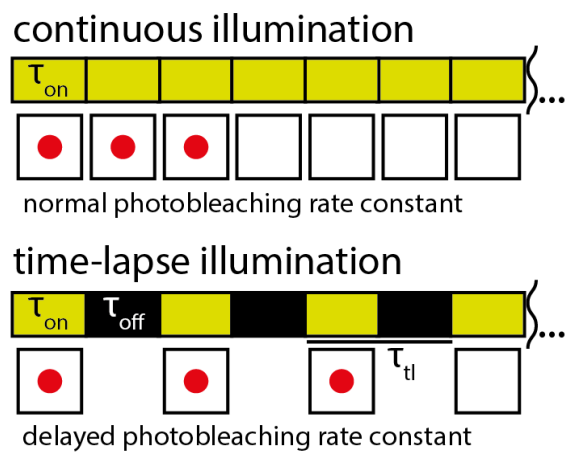

**Supplementary figure S6**
**Illumination scheme for time-lapse microscopy** Scheme of time-lapse microscopy to determine absolute DNA residence times. The red sphere indicates a detected mEos2 molecule. $\tau_{on}$: time laser is on, $\tau_{off}$: time laser is off, $\tau_{tl}$: time-lapse time.

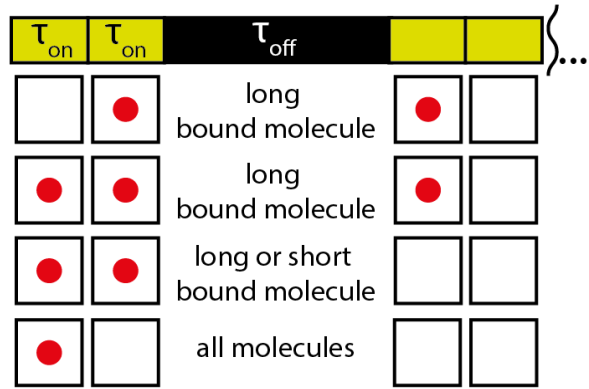

**Supplementary figure S7**
**Illumination scheme for interlaced time-lapse microscopy (ITM)** Scheme of interlaced time-lapse microscopy to determine the ratio of DNA residence time populations. $\tau_{on}$: time laser is on, $\tau_{off}$: time laser is off

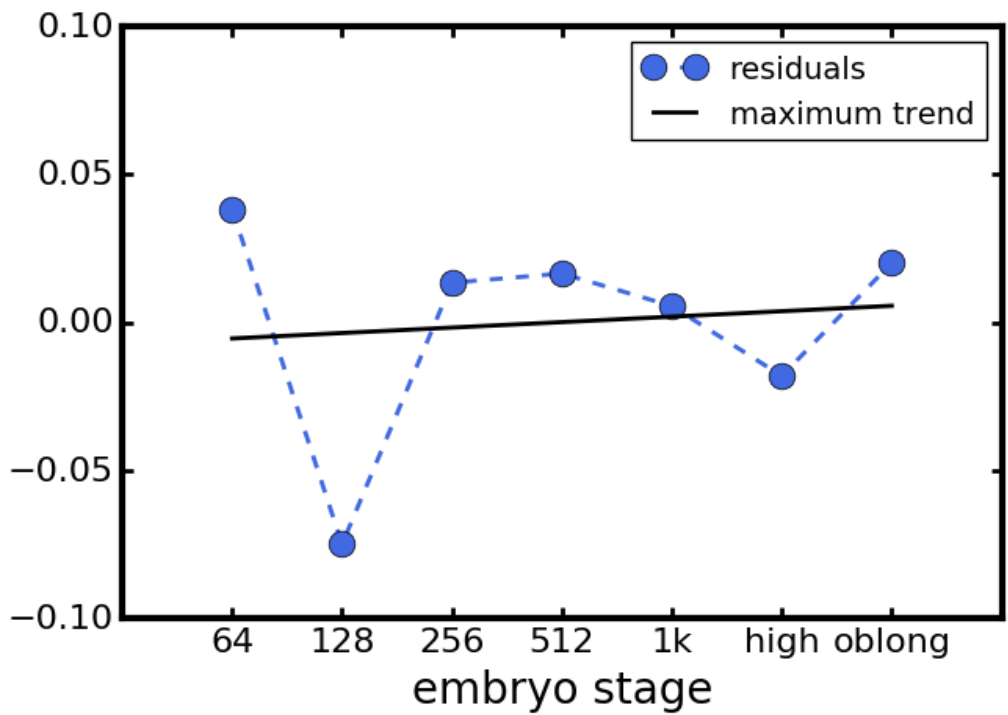

**Supplementary figure S8**
**Residual error between data and model** The residual error between the model and our data is low and distributed around 0 leaving little room for additional trends. The black line indicates a linear fit to the residuals to detect hidden trends. The dashed line is a guide to the eye.

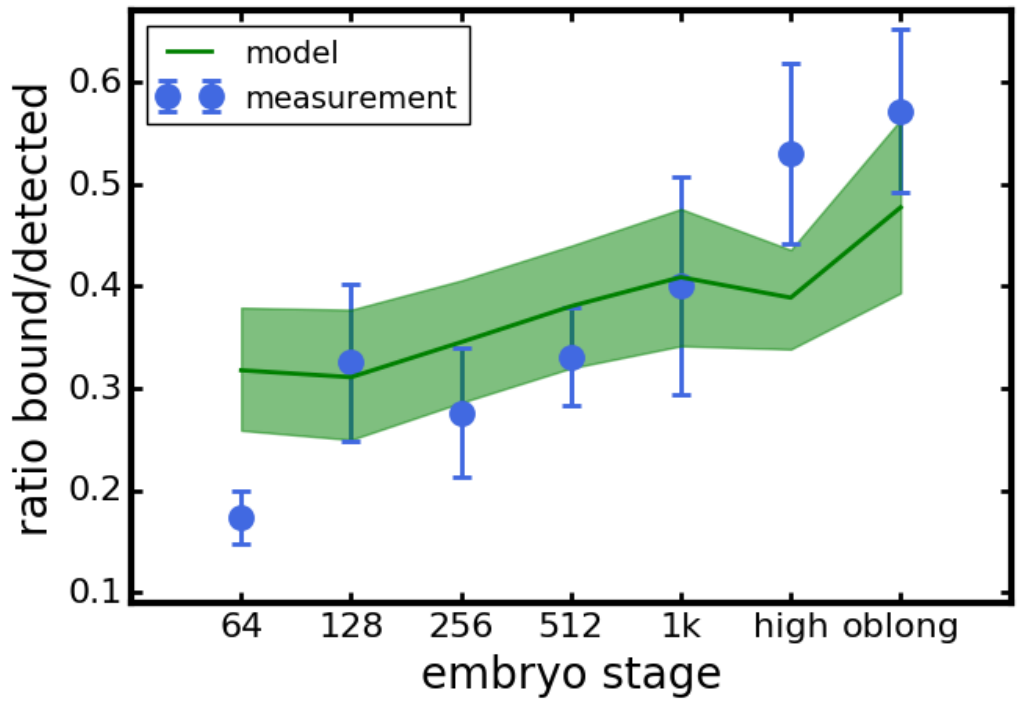

**Supplementary figure S9**
**Fraction of bound molecules in the presence of TSA** Ratio of bound to detected molecules after TSA injection in the 1-cell stage. The model takes into account the binding sites and TBP concentration measurements in TSA injected embryos. Data are represented as mean ± sem (312190 molecules in 3 embryos ).

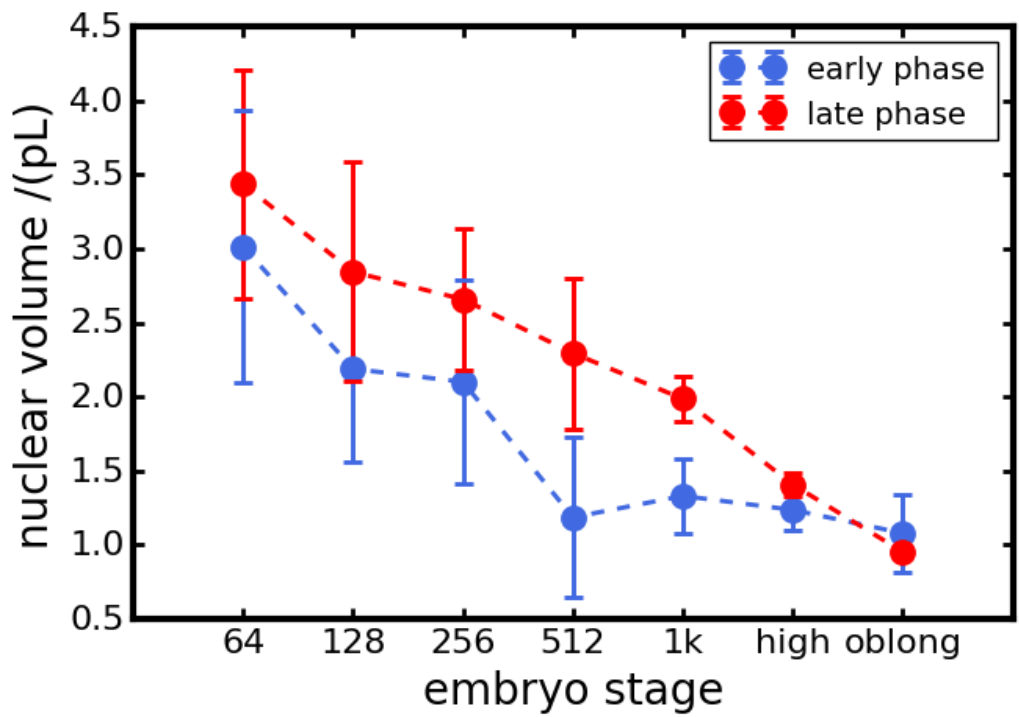

**Supplementary figure S10**
**Nuclear volumes at early and late cell-cycle phase** Nuclear volume increases during the cell cycle. Data are represented as mean ± sem ( 4 embryos ).

# 2 Supplementary Movies

**Supplementary movie S1**
**Imaging of mEos2-TBP inside the zebrafish embryo at the 64-cell stage.** Video was taken with 50ms integration time at a framerate of 20 fps. The surface of the animal cap and the outline of the nucleus are indicated (white lines).

**Supplementary movie S2**
**Imaging of mEos2-TBP in the oblong stage at the periphery of the embryo.** Video was taken with 50ms integration time at a framerate of 20 fps. The surface of the animal cap and the outline of the nucleus are indicated (white lines).

# 3 Supplementary Text

## 3.1 Interlaced Time-Lapse Microscopy (ITM)

Consider an ensemble of fluorescent molecules separated into two different classes of binding affinities to DNA (long and short binding) where the fraction of long or short bound molecules or the binding times undergo a change over time.

### 3.1.1 Optimized temporal illumination scheme and classification of molecules

To optimize the time needed to collect a sufficient amount of information to sense changes in fraction or binding time, the number of time-lapse conditions is reduced to two. Two frames are recorded without pause and the illumination is interrupted for a certain dark time $t_d$, before two new frames are recorded (see Supplementary text figure ST 1). Only trajectories of molecules are considered that move less than 165 nm within two adjacent frames. Detected molecules are then separated into trajectories detected only in two frames without a dark time and trajectories surviving at least one dark time. By carefully choosing the dark time, only molecules of the long bound fraction possibly survive the pause time, whereas in the set of trajectories without a dark time molecules of both fractions can be found. By dividing the number of molecules in both classes a concentration-independent measure $R$ for the DNA affinity of the ensemble can be defined.

$$R = \frac{\text{\# trajectories with dark time}}{\text{\# trajectories with dark time} + \text{\# trajectories w/o dark time}} \quad (1)$$

R, however, does not directly reflect the percentage of all specifically bound molecules, since it is biased by i) the portion of specifically bound molecules that bleach before surviving a dark time and ii) the finite probability of a specifically bound molecule to leave its binding site even before having survived a pause. We therefore developed a correction formula to take these two aspects into account.

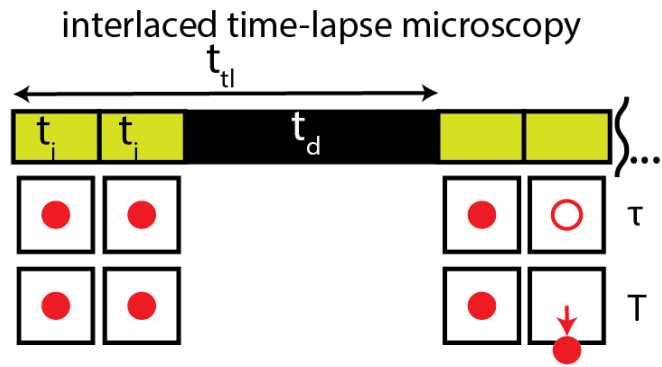

**Supplementary text figure ST 1. Illumination scheme in ITM.** Two successive illumination times $t_i$ are followed by a dark time $t_d$ over a total duration $t_{tl}$. A fluorescent spot trajectory (red dot) identified as a bound molecule can be terminated by two events: by photo-bleaching after time $\tau$ (empty red circle) or unbinding of the molecule after time T (red filled dot leaving the specified area)

### 3.1.2 Linking the ratio stable to all bound molecules to quantitative values

#### 3.1.2.1 Binding time distribution of fluorescent molecules in ITM

If a fluorescent molecule binding to DNA is observed at time $t = 0$, the signal can be lost due to either bleaching or unbinding from DNA. Be $T$ a random variable describing the time the molecule stays bound to DNA and be $\tau$ a random variable describing the time the molecule stays fluorescent then the observed time is given by $\theta = \min(T, \tau)$. If $T$ and $\tau$ are independent the probability for observing a binding time $\theta$ longer than a time $t$ is given by:

$$\Pi(t) = P(\theta > t) = P(\min(T, \tau) > t) = P(\{T > t\} \cap \{\tau > t\}) = P(T > t) \cdot P(\tau > t) \quad (2)$$

For the binding time T, the probability density $f_T$ and the survival function $P(T > t)$ for the two-species ensemble with percentage B and off-rates $k_1$ and $k_2$, is given by:

$$f_T(t) = B k_1 \exp(-k_1 t) + (1 - B) k_2 \exp(-k_2 t) \quad (3)$$

$$P(T > t) = \int_t^\infty f_T(t') dt' = B \exp(-k_1 t) + (1 - B) \exp(-k_2 t) \quad (4)$$

For the fluorescence time $\tau$, the probability of bleaching is not constant in time but zero during the dark time. Therefore the probability density for the fluorescence time $f_\tau$ for a trajectory starting with two illuminated frames of duration $t_i$ each followed by a dark time of duration $t_d$ yielding a total cycle time of $t_{tl} = 2t_i + t_d$ is modeled by the following formula 5:

$$f_\tau(t) = \begin{cases} k_b \exp\left(-k_b \left\lfloor \frac{t}{t_{tl}} \right\rfloor \cdot 2 t_i - k_b \left(t - \left\lfloor \frac{t}{t_{tl}} \right\rfloor \cdot t_{tl}\right)\right) & \left(t - \left\lfloor \frac{t}{t_{tl}} \right\rfloor \cdot t_{tl}\right) < 2 t_i \\ 0 & \left(t - \left\lfloor \frac{t}{t_{tl}} \right\rfloor \cdot t_{tl}\right) \geq 2 t_i \end{cases} \quad (5)$$

Accordingly for the survival function $P(\tau > t)$:

$$P(\tau > t) = \begin{cases} \exp\left(-k_b \left\lfloor \frac{t}{t_{tl}} \right\rfloor \cdot 2 t_i - k_b \left(t - \left\lfloor \frac{t}{t_{tl}} \right\rfloor \cdot t_{tl}\right)\right) & \left(t - \left\lfloor \frac{t}{t_{tl}} \right\rfloor \cdot t_{tl}\right) < 2 t_i \\ \exp\left(-k_b \left\lfloor \frac{t}{t_{tl}} \right\rfloor \cdot 2 t_i\right) & \left(t - \left\lfloor \frac{t}{t_{tl}} \right\rfloor \cdot t_{tl}\right) \geq 2 t_i \end{cases} \quad (6)$$

Supplementary text figure ST 2 shows the survival function $\Pi(t)$ of the observation time $\theta$ for the experimentally found parameters $k_1 = 0.15/s$, $k_2 = 3.92/s$, $k_b = 7.30/s$ and an exemplary value of $B = 1$ (see Supplementary text figure ST 3 A).

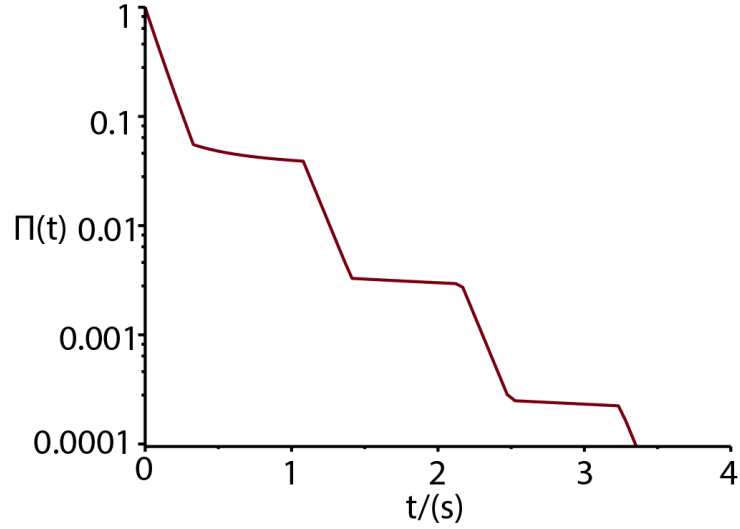

**Supplementary text figure ST 2.** Survival probability of the fluorescent on-state over time in interlaced time-lapse microscopy

**3.1.2.2 Different starting points of the trajectories**   Possible starting points for trajectories are the first or the second frame within the illumination cycle. The fractions of trajectories expected to start at each frame can be calculated by the probability to observe a molecule later than at a certain time point $t_1$, if it has bound at a time point $t_0$. Namely for an exponential model:

$$P(t > t_1 | t_0) = P(t > t_1 - t_0) \tag{7}$$

The probability for a trajectory starting in either the first or the second frame can therefore be calculated by the likelihood of a molecule having bound in either the dark time plus 1 frame time, or within one frame. Although the latter case seems to be highly unlikely, it is not. The high photobleaching rate and the high off-rate make it more likely for a molecule to have bound in temporal vicinity to the observation time point, than a long time before.
So two cases have to be distinguished:

- First-frame trajectories: these molecules have time from the end of the previous frame during the dark time plus the half of the illumination cycle to bind. Assuming a constant off-rate, we can give a normalized measure for the probability in terms of $\Pi$.

$$p'_{1f} = \frac{\Pi(t_i) - \Pi(t_{tl})}{\Pi(t_i)} \tag{8}$$

- Second-frame trajectories: Here the molecule has to bind within one frame time $t_f$ in the illumination cycle. The equivalent probability is therefore given by:

$$p'_{2f} = 1 - \Pi(t_f) \tag{9}$$

Furthermore the survival function $\Pi(t)$ has to be re-normalized in this case as the first illumination cycle is missing. This leads to the survival function $\Pi_1(t)$ that is given by:

$$\Pi_1(t) = P(\theta > t | \theta > t_i) = \frac{P(\theta > t)}{P(\theta > t_i)} = \frac{\Pi(t)}{\Pi(t_i)} \quad (10)$$

As one of both cases must be true, the probability has to be normalized:

$$p_{2f} = \frac{p'_{2f}}{p'_{2f} + p'_{1f}} \quad (11)$$

$$p_{1f} = \frac{p'_{1f}}{p'_{2f} + p'_{1f}} \quad (12)$$

By combining the above equations, we find for the probabilistic weight of all molecules surviving at least one dark time:

$$P(\theta > t_p) = \frac{p_{1f}\Pi(t_{tl}) + p_{2f}\Pi_1(t_{tl})}{p_{1f}\Pi(t_i) + p_{2f}\Pi_1(t_{tl})} \quad (13)$$

This ratio is a function of all decay rates $k_1$, $k_2$, $k_b$ of the process as well as of all time settings $t_i$, $t_{tl}$, $t_f$ and the fraction B, as defined in equation 3. By inverting the equation one can find the correction factor for the measured ratio R, as defined in equation 1 to extract the true fraction B.

### 3.1.3 Correction for the percentage of all bound molecules

Using ITM, it is also possible to compare the total number of all trajectories showing binding at the same spot for two frames with the total number of localizations in the movie. This yields a measure for the percentage of bound and freely diffusing molecules. In our experiment, the number of free molecules, however, will be underestimated as an integration time of 50ms is not able to detect all freely diffusing molecules. Also, the majority of bound molecules is bleaching or unbinding within the first frame and thus counted as unbound molecules. To correct for this we can calculate the percentage of bleached or unbinding molecules within a first frame by computing the survival probability for the molecules over the first frame :

$$P(\theta > t_i) = p_{1f}\Pi(t_i) + p_{2f}\Pi_1(t_{tl}) \quad (14)$$

Thus the probability for mistakenly being counted as non-bound molecule is given by:

$$P(\theta \leq t) = 1 - p_{1f}\Pi(t_i) + p_{2f}\Pi_1(t_{tl}) \quad (15)$$

Let the ratio F of bound to all molecules be calculated by division by the number of all detected molecules:

$$F = \frac{\text{number of bound molecules}}{\text{number of all detections}} \quad (16)$$

To describe the true fraction D of bound molecules, F has to be corrected as follows:

$$D = F \cdot \frac{1}{P(\theta > t_i)} \quad (17)$$

The probability therefore is a function of the off-rates, the bleaching rate and the fraction of long and short bound species obtained in the previously described step. An example for a correction curve is shown in Supplementary text figure ST 3 B.

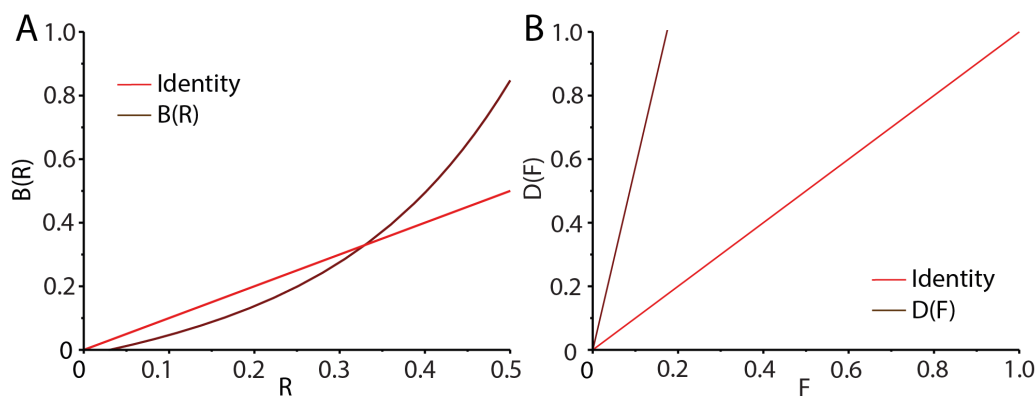

**Supplementary text figure ST 3. Correction of parameters extracted by ITM.** Numbers for stable or short bound molecules obtained by ITM need correction, as photobleaching of the mEos2 fluorophore and unbinding of stable bound TBP molecules before the termination of a dark time biases the distribution towards short binding or non-binding events. (**A**) The ratio of stable to all bound molecules, B, can be under- or overestimated by the ratio R of molecules surviving at least one dark time to all molecules surviving at least two frames, that is experimentally determined by ITM and needs to be corrected. As an example, values of the rate constants for long ($k_1$) and short ($k_2$) binding and bleaching ($k_b$) extracted by time-lapse microscopy are $k_1 = 0.15/s$, $k_2 = 3.92/s$, $k_b = 7.30/s$. The time settings for the illumination time ($t_i$) and the total time-lapse duration $t_{tl}$ are $t_i = 0.165s$, $t_{tl} = 1.08s$. Evaluating and inverting equation 13 for these values using the computer algebra program MAPLE 17 and plotting B as a function of R is shown. In red the ideal one-to-one correction is shown. Below a ratio of R of approx. 32%, B is overestimated. Above a ratio of R of approx. 32% B is underestimated. (**B**) The percentage of all bound molecules compared to all molecules present in the nucleus, D, can be addressed by ITM. According to Supplementary Information section 3.1.3 we can give a correction for the ratio F of all molecules surviving at least two frames compared to all detected molecules. Here as well, molecules bleaching within the first frame will falsely be counted as non-bound molecules. As an example, we plot here the correction curve that is obtained using equation 17 and the values $k_1 = 0.15/s$, $k_2 = 3.92/s$, $k_b = 7.30/s$ and a exemplary value of 0.3 for B.

## 3.2 Physical chemistry of DNA-TBP complex formation in the nuclear confinement

Here, we implement a model that describes the formation of complexes between TBP and specific or unspecific binding sites on DNA in a cell nucleus. The kinetic scheme is depicted in Supplementary Text Figure ST 4.

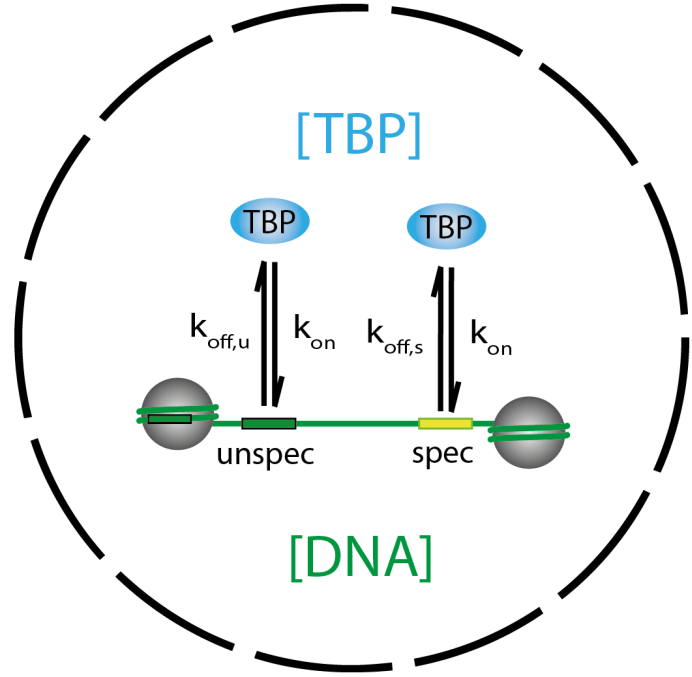

**Supplementary Text Figure ST 4. Kinetic scheme of the concentration model.** [TBP]: concentration of TATA binding protein, [DNA]: concentration of DNA, $k_{on}$: association rate constant, $k_{off,s}$, $k_{off,u}$: TBP dissociation rate constant from specific (s) and unspecific (u) target sequences.

### 3.2.1 Exact solution for TBP binding to two types of binding sites

For the mathematical description of the situation we utilized the law of mass action. We extended it to two binding sites. Be the TBP-concentration denoted by $[T]$, the concentration of unspecific DNA binding sites by $[D_u]$ and the concentration of specific DNA binding sites by $[D_s]$ The model implicates the following equilibria:

$$D_u + T \xrightleftharpoons{K_u} D_u T \quad \text{with} \quad K_u = \frac{[D_u][T]}{[D_u T]} \tag{18}$$

$$D_s + T \xrightleftharpoons{K_s} D_s T \quad \text{with} \quad K_s = \frac{[D_s][T]}{[D_s T]} \tag{19}$$

The total number of each species is given by the sum of free and bound molecules:

$$[T]_0 = [T] + [D_u T] + [D_s T] \tag{20}$$

$$[D_u]_0 = [D_u] + [D_uT], \quad [D_s]_0 = [D_s] + [D_sT] \tag{21}$$

Let $[D] = [D_u] + [D_s]$ denote the effective concentration of DNA binding sites. Then the overall reaction schemes are reduced to:

$$D + T \xrightleftharpoons{K} DT \quad \text{with} \quad K = \frac{[D][T]}{[DT]} \tag{22}$$

The dissociation constant for overall TBP binding to any DNA site can be rewritten as follows:

$$K = \frac{[D][T]}{[DT]} = \frac{([D_u] + [D_s])[T]}{([D_uT] + [D_sT])} = BK_s + (1-B)K_u \tag{23}$$

With the ratio of concentrations of specifically and unspecifically bound molecules denoted by

$$B = \frac{[D_sT]}{[D_uT] + [D_sT]} \tag{24}$$

Assuming the same on-rate $k_{on}$ for both specific and unspecific binding sites we find:

$$K = BK_s + (1-B)K_u = \frac{Bk_s + (1-B)k_u}{k_{on}} \tag{25}$$

In words, this means that the effective dissociation constant is given as the weighted average of the single dissociation constants of long and short bound molecules.

To determine the values of B, we measured the number of specifically bound molecules in the observation volume $V_{ob}$ normalized by the number of all bound molecules, R:

$$R = \frac{[D_sT]V_{ob}}{[D_sT]V_{ob} + [D_uT]V_{ob}} = \frac{[D_sT]}{[D_sT] + [D_uT]} = B \tag{26}$$

R thus links the mathematical description of the model to our measurement.

### 3.2.2 Complex formation depends on actual concentrations of the species

In our single molecule experiments approx. 1000 molecules in the nucleus are searching for potentially millions of binding sites on DNA. We therefore can assume that the concentration of accessible DNA-binding does not change significantly due to binding of TBP

$$[D] = [D_0] - [DT] \approx [D_0] \tag{27}$$

Rewriting the law of mass action with this approximation yields a measure $Q$ that is independent of the concentration of TBP.

$$Q := \frac{[DT]}{[DT] + [T]} = \left(1 + \frac{[T]}{[DT]}\right)^{-1} = \left(1 + \frac{K}{[D_0]}\right)^{-1} \tag{28}$$

Since Q is independent of the concentration of TBP, Q should be the same for endogenous TBP and mEos2-TBP that we are actually measuring. mEos2-TBP therefore acts as a true probe for the ratio Q experienced by any comparable transcription factor.

To determine the values of Q, we measured the number of all bound molecules in the observation volume $V_{ob}$ normalized by the number of all detected molecules, $B'$:

$$B' = \frac{[DT]V_{ob}}{[DT]V_{ob} + [T]V_{ob}} = \frac{[DT]}{[DT] + [T]} = Q \tag{29}$$

$B'$ thus links the mathematical description of the model to our measurement.

**3.2.2.1 Concentration changes of TBP molecules in the nucleus**  According to the law of mass-action, the physical entity influencing the effective on-rates of molecules to the DNA is their molar concentration. As RLSM intrinsically yields numbers $N^f$ of fluorescently labeled molecules in the observed volume fraction of the nucleus it is a means to determine actual concentrations.

Be $A_i^s$ the mean cross-section area of the nuclei in the i-th cell cycle, $w$ the width of the light sheet, $N_A$ Avogadro's number and $N_i^f$ the number of detected molecules in the i-th cycle. We found $w = 3.29 \mu m$. Thus for the nuclear concentration of detected molecules in the i-th cycle $[T]_i$ we find:

$$[T]_i = \frac{N_i^f}{A_i^s \cdot w \cdot N_A} \tag{30}$$

Meaning that given a constant sheet geometry the number of detected molecules per unit area is proportional to the actual concentration of the molecules. By calculating relative values of concentrations the constant parameters $N_A$ and $w$ are canceled out.

**3.2.2.2 Concentration changes of DNA binding sites in the nucleus**  Assuming that DNA is unable to leave the nucleus the concentration is controlled by the nuclear volume. To determine the nuclear volume in each stage, the cross-section $A_i$ was measured in the i-th stage and a spherical shape was assumed.

The volume of the nucleus $V_i^n$ for the i-th stage is then given by:

$$V_i^n = \sqrt{\frac{A_i^3}{36\pi}} \tag{31}$$

Given a certain number of DNA binding sites $N_D$ and Avogadro's number $N_A$ the molar concentration can be calculated as:

$$[D_0]_i = \frac{N_D}{N_A \cdot V_i^n} \tag{32}$$

The DNA concentration therefore is highly governed by the volume of the nucleus.

### 3.2.3 The ratio Q depends on the nuclear size

Combining the above information, we find for the ratio $Q_i$ in the i-th stage

$$Q_i := \left(1 + \frac{K_i}{[D_0]_i}\right)^{-1} = \left(1 + \frac{(R_i k_{off,s} + (1-R_i) k_{off,u}) \cdot N_A \cdot V_i^n}{k_{on} \cdot N_D}\right)^{-1} \tag{33}$$

In eq. 33 every entity is measured except $N_D$ that can be treated as fitting parameter to infer the number of accessible binding sites on DNA.

### 3.2.4 Fit of the model to the data

Fitting procedures were performed using the Levenberg-Marquart algorithm for least-squares optimization. Dependent variable is the all bound/ all detected ratio Q. The function optimized to the all-bound/ all-detected ratio Q is given by:

$$f(N_D, k_{on}) = \frac{[DT]}{[DT] + [T]} \tag{34}$$

Boundary conditions and relations between parameters are :

- The on-rate is assumed to be $0.2 \frac{1}{s \cdot \mu M}$
- The number of binding sites on DNA must be larger than 0.

So, by taking the measured parameters into account, the set of parameters in the function is reduced to:

$$f(N_D) = \frac{[DT]}{[DT] + [T]} = Q \tag{35}$$

The independent variable in our measurements is the developmental stage i.

$$f(i \mid N_D) = Q_i \tag{36}$$

To compare the model with the data, we gave a region of confidence by estimating the upper and lower bound of $f(i \mid N_D)$ by adding or subtracting the individual errors to each variable of the function. This function is then optimized to the experimentally measured ratio $Q_i$. As there might be a scaling factor in the number of DNA binding sites, either due to our measurements or the biochemistry of binding-site recognition of mEos2-TBP, we do not treat the number of binding sites as absolute values but as relative ones in comparable measurement (see Figure 3b and Figure 4).

### 3.2.5 Reconstruction of number of long bound molecules

To characterize the influence of the stable binding time and the concentration of TBP molecules on the number of stable TBP-DNA complexes, we reconstructed this number in the i-th stage $S_i$ from our measurement of relative TBP concentration $[T]_i$, the nuclear volume $V_i^n$, the ratio $Q_i$ of bound molecules to all detected molecules and the average ratio $B_i$ of stable bound molecules to all bound molecules.

$$S_i = [T]_i \cdot V_i^n \cdot Q_i \cdot <B_i> \tag{37}$$

We compared this number with the predictions of the model where we replaced the measured ratio $Q_i$ of bound molecules to all detected molecules but kept the measured values for $[T]_i$, $V_i^n$ and $<B_i>$. Thus, the model value $S_i^m$ for long bound molecules is given by

$$S_i^m = f(i \mid N_D) \cdot [T]_i \cdot V_i^n \cdot <B_i> \tag{38}$$

**Influence of TBP concentration on the number of long bound molecules** To assess the influence of TBP concentration on the number of long bound molecules we multiplied the TBP concentration with a factor $\alpha$. We find

$$S_i^m = f(i \mid N_D) \cdot \alpha[T]_i \cdot V_i^n \cdot <B_i> \qquad (39)$$

This tells us that the number of stable bound molecules is directly proportional to the number of present TBP-molecules as long as DNA is not over-saturated. This is highly expected by the simple law of mass action.

**Influence of stable binding time on the number of long bound molecules** In our model, the stable binding time enters twice, first as a part of the average affinity of TBP to DNA and second since it changes the ratio of long to all bound molecules. Only increasing the long binding time (meaning the lower off-rate $k_{off,s}$) changes the dissociation constant (see eq. 25) but is also disproportionately affecting the ratio of long binding molecules $<B_i>$. By combining eq. 18 and 19 we find

$$\frac{[D_s T]}{[D_u T]} = \frac{K_u[D_s]}{K_s[D_u]} \qquad (40)$$

$$\frac{[D_s T]}{[D_u T] + [D_s T]} = \left(\frac{[D_u T]}{[D_s T]} + 1\right)^{-1} = \left(\frac{k_{off,s}[D_u]}{k_{off,u}[D_s]} + 1\right)^{-1} \qquad (41)$$

For the average ratio of long to all bound molecules $\bar{B}$ as a function of the specific off-rate $k_{off,s}$ this yields:

$$\bar{B}(k_{off,s}) = \left(\frac{k_{off,s}}{3.92/s} \cdot \beta + 1\right)^{-1} \qquad (42)$$

The parameter $\beta$ contains the measured value for the term $\frac{[D_u]}{k_{off,u}[D_s]}$ that is treated as constant and is given by

$$\beta = \left(\frac{1}{<B_i>} - 1\right) \cdot \frac{3.92/s}{0.15/s} \qquad (43)$$